\documentclass[onecolumn]{article}
\usepackage{amsfonts}

\usepackage{graphicx}
\usepackage{amsmath}

\input{tcilatex}

\begin{document}

\title{Optimal Quantum Feedback Control for Canonical Observables }
\author{John Gough\thanks{%
john.gough@ntu.ac.uk} \\
%EndAName
Department of Computing \& Mathematics\\
Nottingham-Trent University, Burton Street,\\
Nottingham NG1\ 4BU, United Kingdom.}
\date{}
\maketitle

\begin{abstract}
We show that the stochastic Schr\"{o}dinger equation for the filtered state
of a system, with linear free dynamics, undergoing continual non-demolition
measurement or either position or momentum, or both together, can be solved
explicitly within a class of Gaussian states which we call extended coherent
states. The asymptotic limit yields a class of relaxed states which we
describe explicitly. Bellman's principle is then applied directly to optimal
feedback control of such dynamical systems and the Hamilton Jacobi Bellman
equation for the minimum cost is derived. The situation of quadratic
performance criteria is treated as the important special case and solved
exactly for the class of relaxed states. 

PACS numbers: 07.55.Ge, 42.50.Lc, 03.65.Ta, 05.45.Mt
\end{abstract}

\section{Introduction}

Quantum noise was originally developed to model irreversible quantum
dynamical systems, where it played an external and secondary role, however,
the realization that it could be measured and the results used to influence
the system evolution has had a profound effect on its physical status \cite
{Belavkin89},\cite{CollettGardiner},\cite{Belavkin99}. The great leap
forward since then has been made by experimentalists who have made the
practical implementation of quantum state estimation and adaptive feedback
control a reality. With this, has come new problems that have received
intense interest in the physics community [4-10].

In this paper, we wish to treat the problem of how to describe the quantum
evolution of a system with linear free dynamics when we perform
non-demolition measurements of, typically both, canonical position and
momentum. The problem where position measurements only are made has been of
historical importance. In this situation, the model is the one considered by
Ghirardi, Rimini and Weber \cite{GRW}, who also obtained the asymptotic form
for the state. The asymptotic solution, with explicit reference to the
stochastic Schr\"{o}dinger equation within the It\^{o} formulation, was
first given by Di\'{o}si \cite{Diosi(a)}, see also Belavkin and Staszewski 
\cite{BS}. Essentially, the solution to the stochastic Schr\"{o}dinger
equation could be understood as an randomly parameterized Gaussian state.
The parameters being mean position, mean momentum and a complex inverse
variance. We shall show that the same class of states, which we term
extended coherent states, suffice for the stochastic Schr\"{o}dinger
equation describing simultaneous monitoring of position and momentum.

The problem of optimal quantum feedback control can then be tackled at this
point. Bellman equations have been derived previously for the optimal cost
of controlling a qubit system \cite{BoutenEdwardsBelavkin}. In fact, the
general problem can be understood as a classical control problem on the
space of quantum states \cite{GSB} if one exploits the separation of quantum
estimation component from the control component: here we may construct a,
typically infinite dimensional, Hamilton Jacobi Bellman theory and are then
faced with the problem of finding a sufficient parameterization of states
for particular situation. In the case of non-demolition position and
momentum measurements, we have that the extended coherent states offer a
sufficient parameterization. The quadratic performance problem is the
important special case and has been treated by Doherty and Jacobs \cite{DJ}\
for feedback from measuring one quadrature of a Bosonic mode. We show that
this problem is solvable when both canonical observables are measured.

\subsection{Stochastic Schr\"{o}dinger Equation}

Consider a quantum system evolving with free Hamiltonian $H$ while
undergoing continual diffusive interaction with several independent
apparatuses, each coupling to the system in a Markovian manner with coupling
operator $L_{j}$ for the $j$-th apparatus. (The $\left\{ L_{j}\right\} $ do
not generally need to be either commuting or self-adjoint.) The state, $\psi
_{t}$, of the system continually updated using the output of the
apparatuses, will then satisfy a stochastic Schr\"{o}dinger equation of the
type \cite{Diosi},\cite{Belavkin99},\cite{GutaBoutenMaassen}, 
\begin{eqnarray}
\left| d\psi _{t}\right\rangle &=&\frac{1}{i\hbar }H\left| \psi
_{t}\right\rangle \,dt-\frac{1}{2}\sum_{j}\left( L_{j}^{\dagger
}L_{j}-2\lambda _{j}\left( t\right) L_{j}+\lambda _{j}^{2}\left( t\right)
\right) \left| \psi _{t}\right\rangle \,dt  \notag \\
&&+\sum_{j}\left( L_{j}-\lambda _{j}\left( t\right) \right) \left| \psi
_{t}\right\rangle \,dW_{t}^{\left( j\right) }.  \label{SSE}
\end{eqnarray}
where $\lambda _{j}\left( t\right) =\func{Re}\left\langle \psi
_{t}|L_{j}\,\psi _{t}\right\rangle $ and $\left\{ W^{\left( j\right)
}\right\} $ is a multi-dimensional Wiener process with $dW_{t}^{\left(
j\right) }dW_{t}^{\left( k\right) }=\delta _{jk}dt$. This equation was first
postulated in the context of filtering by Belavkin where the apparatuses are
separate Bose fields and the $W_{t}^{\left( j\right) }$ are innovations
processes obtained by de-trending the output processes.

The stochastic Schr\"{o}dinger equation for measurement of canonically
conjugate observables, $\hat{q}$ and $\hat{p}$, has been derived from first
principles by Scott and Milburn \cite{SM}. They considered a discrete time
model with simultaneous measurement of position and momentum by separate
apparatuses, and considered the continuous time limit of progressively more
imprecise and frequent measurements. Taking $L_{1}=\sqrt{\dfrac{\kappa }{2}}%
\hat{q}$ and $L_{2}=\sqrt{\dfrac{\tilde{\kappa}}{2}}\hat{p}$ and denoting
the innovations by $W_{t}^{\left( 1\right) }=W_{t}$ and $W_{t}^{\left(
2\right) }=\tilde{W}_{t}$, their particular stochastic Schr\"{o}dinger
equation reads as 
\begin{eqnarray}
\left| d\psi _{t}\right\rangle &=&\left( \frac{1}{i\hbar }H-\frac{\kappa }{4}%
\left( \hat{q}-\left\langle \hat{q}\right\rangle _{t}\right) ^{2}-\frac{%
\tilde{\kappa}}{4}\left( \hat{p}-\left\langle \hat{p}\right\rangle
_{t}\right) ^{2}\right) \left| \psi _{t}\right\rangle \,dt  \notag \\
&&+\sqrt{\frac{\kappa }{2}}\left( \hat{q}-\left\langle \hat{q}\right\rangle
_{t}\right) \left| \psi _{t}\right\rangle \,dW_{t}+\sqrt{\frac{\tilde{\kappa}%
}{2}}\left( \hat{p}-\left\langle \hat{p}\right\rangle _{t}\right) \left|
\psi _{t}\right\rangle \,d\tilde{W}_{t}.  \label{SSEqp}
\end{eqnarray}

The equation involves the expectations $\left\langle \hat{q}\right\rangle
_{t}=\left\langle \psi _{t}|\hat{q}\,\psi _{t}\right\rangle $ and $%
\left\langle \hat{p}\right\rangle _{t}=\left\langle \psi _{t}|\hat{p}\,\psi
_{t}\right\rangle $ and is therefore non-linear in the state $\psi _{t}$.
Here the constants $\kappa $ m$^{-2}$s$^{-1}$\ and $\tilde{\kappa}$ N$^{-2}$s%
$^{-3}$ are positive and describe the measurement strength for the two
apparatuses. In general, $\kappa $ and $\tilde{\kappa}$ has units of inverse
variance of position, respectively momentum, per unit time. In \cite{GS},
the limiting procedure was revisited and, as an alternative to increasingly
imprecise measurements, one could use increasingly weak interaction between
the apparatuses and the system. The scaling between the imprecision of
measurement, or weakness of interaction with the apparatus, and the rate at
which the discrete measurements is made must be such as to allow a general
central limit effect to take place. In principle, it is possible, to set up
the apparatuses to obtain desired values of $\kappa $ and $\tilde{\kappa}$.

The purpose of \cite{SM} was to consider nonlinear dynamics, however, we
shall only deal with quadratic Hamiltonians of the type $H=H\left(
f,v\right) $%
\begin{equation}
H=\frac{1}{2m}\hat{p}^{2}+\frac{1}{2}\hbar \mu \hat{q}^{2}-f\hat{q}+v\hat{p}.
\label{H}
\end{equation}
Here $f$ and $v$ are external fields which will later be replaced with
control functions. We shall show that it is possible to find a general
solution for the stochastic state $\psi _{t}$, with initial condition being
that we start in a coherent state, realized as a random wave function taking
values in a special class of wave functions, termed extended coherent states.

\section{Extended Coherent States}

Let $L^{2}\left( \mathbb{R}\right) $ be the Hilbert space of square
integrable functions of position coordinate $x$ with standard
Schr\"{o}dinger representation of the canonical observables $\hat{q}$ and $%
\hat{p}$. By an \textit{extended coherent state}, we mean a\ wave function $%
\psi \left( \bar{q},\bar{p},\eta \right) $, parameterized by real numbers $%
\bar{q},\bar{p}$ and a complex number $\eta =\eta ^{\prime }+i\eta ^{\prime
\prime } $ where $\eta ^{\prime }>0$, taking the form 
\begin{equation}
\left\langle x|\psi \left( \bar{q},\bar{p},\eta \right) \right\rangle
=\left( \frac{\eta ^{\prime }}{2\pi }\right) ^{1/4}\exp \left\{ -\frac{\eta 
}{4}\left( x-\bar{q}\right) ^{2}+i\frac{\bar{p}}{\hbar }x\right\} .
\end{equation}
When $\eta $ is real $\left( \eta ^{\prime \prime }=0\right) $, the vectors
are just the well-known coherent states \cite{Louisell} The distribution of
the canonical variables in extended coherent state $\psi \left( \bar{q},\bar{%
p},\eta \right) $ is Gaussian with characteristic function 
\begin{equation}
\left\langle \exp \left\{ ir\hat{q}+is\hat{p}\right\} \right\rangle _{\bar{q}%
,\bar{p},\eta }=\exp \left\{ ir\bar{q}+is\bar{p}-\frac{1}{2}\left(
C_{qq}r^{2}+2C_{qp}rs+C_{pp}s^{2}\right) \right\} ,  \label{Weyl expectation}
\end{equation}
where 
\begin{equation}
C_{qq}=\frac{1}{\eta ^{\prime }},\;C_{qp}=-\frac{\hbar \eta ^{\prime \prime }%
}{2\eta ^{\prime }},\;C_{pp}=\frac{\hbar ^{2}}{4}\left( \eta ^{\prime }+%
\frac{\eta ^{\prime \prime 2}}{\eta ^{\prime }}\right) .  \label{covariances}
\end{equation}
The mean values of the position and the momentum in an extended coherent
state are evidently $\left\langle \hat{q}\right\rangle =\bar{q}$ and $%
\left\langle \hat{p}\right\rangle =p$ respectively. We have that $C_{qq}$ is
the variance of $\hat{q}$, $C_{pp}$ is the variance of $\hat{p}$, while $%
C_{qp}=\frac{1}{2}\left\langle \hat{q}\hat{p}+\hat{p}\hat{q}\right\rangle
-\left\langle \hat{p}\right\rangle \left\langle \hat{q}\right\rangle $ is
the covariance of $\ \hat{q}$ and $\hat{p}$.

\subsection{Derivation of the Characteristic Function}

To establish $\left( \ref{Weyl expectation}\right) $, let us first recall
that coherent states may be constructed from creation/annihilation operators 
$a^{\pm }=\dfrac{1}{2}\sqrt{\eta ^{\prime }}\hat{q}\pm \dfrac{1}{i\hbar 
\sqrt{\eta ^{\prime }}}\hat{p}$ by identifying $\psi \left( \bar{q},\bar{p}%
,\eta ^{\prime }\right) $ as the eigenstate of $a^{-}$ with eigenvalue $%
\alpha =\dfrac{1}{2}\sqrt{\eta ^{\prime }}\bar{q}-\dfrac{1}{i\hbar \sqrt{%
\eta ^{\prime }}}\bar{p}$. In particular, if $\Omega $ denotes the
zero-eigenstate of $a^{-}$ then 
\begin{equation*}
\psi \left( \bar{q},\bar{p},\eta ^{\prime }\right) =D_{\alpha }\,\Omega
\end{equation*}
where $D_{\alpha }=\exp \left\{ \alpha a^{+}-\alpha ^{\ast }a^{-}\right\} $
is a Weyl displacement unitary. Next observe that we may obtain extended
coherent states from coherent states by the simple application of a unitary
transformation: 
\begin{equation*}
\psi \left( \bar{q},\bar{p},\eta ^{\prime }+i\eta ^{\prime \prime }\right)
\equiv V\;\psi \left( \bar{q},\bar{p},\eta ^{\prime }\right)
\end{equation*}
with $V=\exp \left\{ -\frac{i}{4}\eta ^{\prime \prime }\left( \hat{q}-\bar{q}%
\right) ^{2}\right\} $. (This transformation is, in fact, linear canonical.)
We may introduce new canonical variables $\hat{q}^{\prime }$ and $\hat{p}%
^{\prime }$ by $\hat{q}^{\prime }=V^{\dagger }\hat{q}V\equiv \hat{q}$ and $%
\hat{p}^{\prime }=V^{\dagger }\hat{p}V=\hat{p}-\frac{1}{2}\hbar \eta
^{\prime \prime }\left( \hat{q}-\bar{q}\right) $. We note that $\exp \left\{
ir\hat{q}+is\hat{p}\right\} =D_{z}$ where $z=-\dfrac{1}{2}\hbar \sqrt{\eta
^{\prime }}s+i\dfrac{1}{\sqrt{\eta ^{\prime }}}r$ and 
\begin{equation*}
V^{\dagger }D_{z}V=\exp \left\{ ir\hat{q}^{\prime }+is\hat{p}^{\prime
}\right\} =D_{w}\,e^{\frac{1}{2}i\hbar \eta ^{\prime \prime }\bar{q}s}
\end{equation*}
where $w=-\dfrac{1}{2}\hbar \sqrt{\eta ^{\prime }}s+i\dfrac{1}{\sqrt{\eta
^{\prime }}}\left( r-\frac{1}{2}\hbar \eta ^{\prime \prime }s\right) $.
Using well-known properties for Weyl displacement operators \cite{Louisell}
and their $\Omega $-state averages, we find 
\begin{eqnarray*}
\left\langle \exp \left\{ ir\hat{q}+is\hat{p}\right\} \right\rangle _{\bar{q}%
,\bar{p},\eta } &=&\left\langle \Omega |D_{\alpha }^{\dagger }V^{\dagger
}D_{z}VD_{\alpha }\,\Omega \right\rangle \\
&=&\left\langle \Omega |D_{\alpha }^{\dagger }D_{w}D_{\alpha }\,\Omega
\right\rangle e^{\frac{1}{2}i\hbar \eta ^{\prime \prime }\bar{q}s} \\
&=&e^{w\alpha ^{\ast }-w^{\ast }\alpha -\frac{1}{2}|w|^{2}}e^{\frac{1}{2}%
i\hbar \eta ^{\prime \prime }\bar{q}s}
\end{eqnarray*}
and substituting in for $\alpha $ and $w$ gives the required result.

\subsection{Weyl Independence}

We say that the canonical variables are Weyl independent for a given state $%
\left\langle \,\cdot \,\right\rangle $, not necessarily pure, if we have the
following factorization 
\begin{equation*}
\left\langle \exp \left\{ ir\hat{q}+is\hat{p}\right\} \right\rangle
=\left\langle \exp \left\{ ir\hat{q}\right\} \right\rangle \,\left\langle
\exp \left\{ is\hat{p}\right\} \right\rangle 
\end{equation*}
for all real $r$ and $s$. If the state possesses moments to all orders, then
Weyl independence means that symmetrically (Weyl) ordered moments factor
according to $\left\langle :f\left( \hat{q}\right) g\left( \hat{p}\right)
:\right\rangle =\left\langle f\left( \hat{q}\right) \right\rangle
\left\langle g\left( \hat{p}\right) \right\rangle $, for all polynomials $f,g
$. By inspection, we see that coherent states leave the canonical variables
Gaussian and Weyl-independent. However, the $\eta ^{\prime \prime }\neq 0$
extended states do not have this Weyl-independence property.

\section{Stochastic Wave Function}

We now return to the equation $\left( \ref{SSEqp}\right) $ for the
conditioned state $\psi _{t}$. Let $\left\langle X\right\rangle
_{t}=\left\langle \psi _{t}\right| X\left| \psi _{t}\right\rangle $, for a
general operator $X$, then we have the following stochastic Ehrenfest
equation

\begin{gather}
d\left\langle X\right\rangle =\left\{ \frac{1}{i\hbar }\left\langle \left[
X,H\right] \right\rangle -\frac{\kappa }{4}\left\langle \left[ \left[ X,\hat{%
q}\right] ,\hat{q}\right] \right\rangle -\frac{\tilde{\kappa}}{4}%
\left\langle \left[ \left[ X,\hat{p}\right] ,\hat{p}\right] \right\rangle
\right\} \,dt  \notag \\
+\sqrt{\frac{\kappa }{2}}\left( \left\langle X\hat{q}+\hat{q}X\right\rangle
-\left\langle \hat{q}\right\rangle \left\langle X\right\rangle \right)
\,dW_{t}+\sqrt{\frac{\tilde{\kappa}}{2}}\left( \left\langle X\hat{p}+\hat{p}%
X\right\rangle -\left\langle \hat{p}\right\rangle \left\langle
X\right\rangle \right) \,d\tilde{W}_{t}.  \label{filter X}
\end{gather}
For $X=\hat{q},\hat{p}$, we find 
\begin{eqnarray}
d\left\langle \hat{q}\right\rangle &=&\left( \frac{1}{m}\left\langle \hat{p}%
\right\rangle +v\right) \,dt+\sqrt{2\kappa }C\left( \hat{q},\hat{q}\right)
\,dW_{t}+\sqrt{2\tilde{\kappa}}C\left( \hat{q},\hat{p}\right) \,d\tilde{W}%
_{t},  \notag \\
d\left\langle \hat{p}\right\rangle &=&\left( -\hbar \mu \left\langle \hat{q}%
\right\rangle +f\right) \,dt+\sqrt{2\kappa }C\left( \hat{q},\hat{p}\right)
\,dW_{t}+\sqrt{2\tilde{\kappa}}C\left( \hat{p},\hat{p}\right) \,d\tilde{W}%
_{t}.  \label{filter q,p}
\end{eqnarray}
where $C\left( \hat{q},\hat{q}\right) =\left\langle \hat{q}^{2}\right\rangle
-\left\langle \hat{q}\right\rangle ^{2},C\left( \hat{p},\hat{p}\right)
=\left\langle \hat{p}^{2}\right\rangle -\left\langle \hat{p}\right\rangle
^{2},$ and $C\left( \hat{q},\hat{p}\right) =\frac{1}{2}\left\langle \hat{p}%
\hat{q}+\hat{q}\hat{p}\right\rangle -\left\langle \hat{p}\right\rangle
\left\langle \hat{q}\right\rangle $. In the following, we wish to
investigate the dynamical evolution of the random state $\psi $ starting
from an initial coherent state. It turns out however that we do not remain
within the class of coherent states: if we did, then $\hat{q}$ and $\hat{p}$
would remain Weyl-independent and, in particular, $C\left( \hat{q},\hat{p}%
\right) $ would vanish, along with the noise term in the $\left\langle \hat{p%
}\right\rangle $-equation of $\left( \ref{filter q,p}\right) $ above and
this would lead to an inconsistent system of equations. Fortunately, it
turns out that it is possible to think of $\psi $ as evolving as a random
state taking values amongst the extended coherent states. Explicitly, we
make the ansatz that the state $\psi _{t}$ takes the form 
\begin{equation}
\psi _{t}=\psi \left( \bar{q}_{t},\bar{p}_{t},\eta _{t}\right)
\label{ansatz}
\end{equation}
where $\bar{q}_{t}$ and $\bar{p}_{t}$ are real-valued diffusion processes
satisfying \ and $\eta _{t}$ is a complex-valued deterministic function. Our
assumption that we start from a coherent state is equivalent to asking that $%
\eta \left( 0\right) =\sigma ^{-2}>0$, with $\sigma $ having the
interpretation as the initial dispersion in position.

We shall now show that $\bar{q}$, $\bar{p}$ satisfy the diffusion equations $%
\left( \ref{filter q,p}\right) $, while $\eta $ satisfies the Riccati
equation 
\begin{equation}
\frac{d}{dt}\eta =2\kappa +i2\mu -\frac{1}{2}\left( \tilde{\kappa}\hbar
^{2}+i\frac{\hbar }{m}\right) \eta ^{2}.  \label{Riccati}
\end{equation}

\subsection{Consistency with the Statistical Evolution}

Let $r,s$ be fixed real parameters and set $D=\exp \left\{ ir\hat{q}+is\hat{p%
}\right\} $. We shall investigate the evolution through the characteristic
function 
\begin{equation*}
G_{t}=\left\langle \psi _{t}\right| D\left| \psi _{t}\right\rangle
=\left\langle D\right\rangle _{t}.
\end{equation*}
Observing that $\left[ D,\hat{q}\right] =\hbar sD,\;\left[ D,\hat{p}\right]
=-\hbar rD$ we find 
\begin{multline*}
dG=\left\{ \frac{ir}{2m}\left\langle \hat{p}D+D\hat{p}\right\rangle -\frac{%
is\hbar \mu }{2}\left\langle \hat{q}D+D\hat{q}\right\rangle +\left( ifs+ivr-%
\frac{\hbar ^{2}\left( \kappa s^{2}+\tilde{\kappa}r^{2}\right) }{4}\right)
G\right\} \,dt \\
+\sqrt{\frac{\kappa }{2}}\left( \left\langle D\hat{q}+\hat{q}D\right\rangle
-\left\langle \hat{q}\right\rangle G\right) \,dW+\sqrt{\frac{\tilde{\kappa}}{%
2}}\left( \left\langle D\hat{p}+\hat{p}D\right\rangle -\left\langle \hat{p}%
\right\rangle G\right) \,d\tilde{W}.
\end{multline*}
The identity $e^{ir\hat{q}+is\hat{p}}=e^{\frac{1}{2}irs\hbar }e^{ir\hat{q}%
}e^{is\hat{p}}=e^{-\frac{1}{2}irs\hbar }e^{is\hat{p}}e^{ir\hat{q}}$ (Baker
Campbell Hausdorff formula) then allows us to compute that 
\begin{equation*}
\left\langle \hat{q}D\right\rangle =e^{\frac{1}{2}irs\hbar }\frac{1}{i}\frac{%
\partial }{\partial r}\left( e^{-\frac{1}{2}irs\hbar }G\right) =\left( \bar{q%
}+i\left( C_{qq}^{2}r+C_{qp}^{2}s\right) +\frac{1}{2}s\hbar \right) G,
\end{equation*}
and likewise 
\begin{eqnarray*}
\left\langle D\hat{q}\right\rangle  &=&\left( \bar{q}+i\left(
C_{qq}r+C_{qp}s\right) -\frac{1}{2}s\hbar \right) G, \\
\left\langle \hat{p}D\right\rangle  &=&\left( \bar{p}+i\left(
C_{qp}r+C_{pp}s\right) +\frac{1}{2}r\hbar \right) G, \\
\left\langle D\hat{p}\right\rangle  &=&\left( \bar{p}+i\left(
C_{qp}r+C_{pp}s\right) -\frac{1}{2}r\hbar \right) G.
\end{eqnarray*}
Hence 
\begin{eqnarray}
dG &=&\frac{ir}{m}\left\{ \bar{p}+i\left( C_{qp}r+C_{pp}s\right) \right\}
G\,dt-is\hbar \mu \left\{ \bar{q}+i\left( C_{qq}r+C_{qp}s\right) \right\}
G\,dt  \notag \\
&&+\left( ifs+ivr-\frac{\kappa \hbar ^{2}s^{2}}{4}-\frac{\tilde{\kappa}\hbar
^{2}r^{2}}{4}\right) G\,dt  \notag \\
&&+i\sqrt{2\kappa }\left( C_{qq}r+C_{qp}s\right) G\,dW+i\sqrt{2\tilde{\kappa}%
}\left( C_{qp}r+C_{pp}s\right) G\,d\tilde{W}.  \label{dG(1)}
\end{eqnarray}
Under our ansatz $\left( \ref{ansatz}\right) $, we should also have, by the
It\^{o} rule, 
\begin{eqnarray}
dG &=&\frac{\partial G}{\partial \bar{q}}d\bar{q}+\frac{\partial G}{\partial 
\bar{p}}d\bar{p}+\frac{1}{2}\frac{\partial ^{2}G}{\partial \bar{q}^{2}}%
\left( d\bar{q}\right) ^{2}+\frac{\partial ^{2}G}{\partial \bar{q}\partial 
\bar{p}}\left( d\bar{q}d\bar{p}\right) +\frac{1}{2}\frac{\partial ^{2}G}{%
\partial \bar{p}^{2}}\left( d\bar{p}\right) ^{2}  \notag \\
&&+\frac{\partial G}{\partial \eta ^{\prime }}d\eta ^{\prime }+\frac{%
\partial G}{\partial \eta ^{\prime \prime }}d\eta ^{\prime \prime }  \notag
\\
&=&irGd\bar{q}+isGd\bar{p}-\frac{1}{2}r^{2}G\left( d\bar{q}\right)
^{2}-rsG\left( d\bar{q}d\bar{p}\right) -\frac{1}{2}s^{2}G\left( d\bar{p}%
\right) ^{2}  \notag \\
&&+\left( \frac{1}{2\eta ^{\prime 2}}r^{2}-\frac{\hbar \eta ^{\prime \prime }%
}{2\eta ^{\prime }}rs-\frac{\hbar ^{2}}{8}\left( 1-\frac{\eta ^{\prime
\prime 2}}{\eta ^{\prime 2}}\right) s^{2}\right) d\eta ^{\prime }  \notag \\
&&+\left( \frac{\hbar }{2\eta ^{\prime }}rs-\frac{1}{4}\frac{\hbar ^{2}\eta
^{\prime \prime }}{\eta ^{\prime }}s^{2}\right) d\eta ^{\prime \prime }.
\label{dG(2)}
\end{eqnarray}
Equating the coefficients of $\left( \ref{dG(1)}\right) $ and $\left( \ref
{dG(2)}\right) $ gives the system of equations 
\begin{eqnarray*}
r &:&d\bar{q}=\left( \dfrac{1}{m}\bar{p}+v\right) \,dt+\sqrt{2\kappa }%
C_{qq}\,dW+\sqrt{2\tilde{\kappa}}C_{qp}\,d\tilde{W}, \\
s &:&d\bar{p}=\left( -\hbar \mu \bar{q}+f\right) \,dt+\sqrt{2\kappa }%
C_{qp}\,dW+\sqrt{2\tilde{\kappa}}C_{pp}\,d\tilde{W}, \\
r^{2} &:&\left( d\bar{q}\right) ^{2}-\dfrac{1}{\eta ^{\prime 2}}d\eta
^{\prime }=\dfrac{1}{m}C_{qp}dt+\dfrac{\tilde{\kappa}\hbar ^{2}}{2}\,dt, \\
s^{2} &:&\left( d\bar{p}\right) ^{2}+\dfrac{\hbar ^{2}}{4}\left( 1-\dfrac{%
\eta ^{\prime \prime 2}}{\eta ^{\prime 2}}\right) \,d\eta ^{\prime }+\frac{1%
}{2}\dfrac{\hbar ^{2}\eta ^{\prime \prime }}{\eta ^{\prime }}\,d\eta
^{\prime \prime }=-2\hbar \mu C_{qp}\,dt+\dfrac{\kappa \hbar ^{2}}{2}\,dt, \\
rs &:&\left( d\bar{q}d\bar{p}\right) +\dfrac{\hbar \eta ^{\prime \prime }}{%
2\eta ^{\prime 2}}\,d\eta ^{\prime }-\dfrac{\hbar }{2\eta ^{\prime }}\,d\eta
^{\prime \prime }=\frac{1}{m}C_{pp}\,dt-\hbar \mu \,C_{qq}dt.
\end{eqnarray*}
The first two of these agree exactly with $\left( \ref{filter q,p}\right) $,
while the next three are entirely consistent with the pair of real equations 
\begin{equation}
\left\{ 
\begin{array}{c}
\dfrac{d}{dt}\eta ^{\prime }=2\kappa +\dfrac{\hbar }{m}\eta ^{\prime }\eta
^{\prime \prime }-\frac{1}{2}\tilde{\kappa}\hbar ^{2}\left( \eta ^{\prime
2}-\eta ^{\prime \prime 2}\right) , \\ 
\\ 
\dfrac{d}{dt}\eta ^{\prime \prime }=2\mu -\dfrac{\hbar }{2m}\left( \eta
^{\prime 2}-\eta ^{\prime \prime 2}\right) -\tilde{\kappa}\hbar ^{2}\eta
^{\prime }\eta ^{\prime \prime }.
\end{array}
\right. 
\end{equation}
Together, they are equivalent to the single complex Riccati equation $\left( 
\ref{Riccati}\right) $.

\subsection{Asymptotic States}

The Riccati equation $\left( \ref{Riccati}\right) $ is to be solved in the
half plane $\eta ^{\prime }>0$ of physical solutions and has the unique,
globally attractive, fixed point 
\begin{equation}
\eta _{\infty }=\frac{2}{\hbar }\sqrt[+]{\frac{\kappa +i\mu }{\tilde{\kappa}+%
\dfrac{i}{m\hbar }}}.
\end{equation}
(Here $\sqrt[+]{\cdot }$ denotes the complex root having positive real part.)

In the case of a harmonic oscillator of frequency $\omega $, we have $\mu =%
\dfrac{m\omega ^{2}}{\hbar }>0$ and we may achieve a coherent state ($\eta
_{\infty }$ real) as the limit state if we tune the measurement strengths
such that $\kappa \equiv m^{2}\omega ^{2}\,\tilde{\kappa}$. In this case, $%
\eta _{\infty }\equiv \dfrac{2m\omega }{\hbar }$, corresponding to a
coherent state with position uncertainty $\sigma _{\infty }=\sqrt{\dfrac{%
\hbar }{2m\omega }}$. Otherwise the limit state will be an extended coherent
state.

We should remark that $\sqrt{\dfrac{\kappa }{\tilde{\kappa}}}$ corresponds
to the squeezing parameter $s$ introduced in \cite{SM} to describe the bias
in favor of the $\hat{q}$ or $\hat{p}$ coupling.

\section{Optimal Quantum Feedback Control}

We fix a terminal time $T>0$ and let $\left\{ f_{t}:0<t<T\right\} $ and $%
\left\{ v_{t}:0<t<T\right\} $ be prescribed functions which we refer to as
control policies. Let $\psi _{t}=\psi \left( \bar{q}_{t},\bar{p}_{t},\eta
_{t}\right) $ be the solution to the stochastic Schr\"{o}dinger equation
with time-dependent free Hamiltonian $H=H\left( f_{t},v_{t}\right) $ and
initial state being an extended state $\psi \left( \bar{q}_{0},\bar{p}%
_{0},\eta _{0}\right) $ at time $t_{0}$ somewhere in the time interval $%
\left[ 0,T\right] $.

We wish to grade the control policies $\left\{ f_{t}\right\} $ and $\left\{
v_{t}\right\} $ over the time interval $\left[ t_{0},T\right] $ and do so by
assigning a cost $J=J\left[ \left\{ f_{t}\right\} ,\left\{ v_{t}\right\}
;t_{0},T;\bar{q}_{0},\bar{p}_{0},\eta _{0}\right] $ taking the general form 
\begin{equation}
J\left[ \left\{ f_{t}\right\} ,\left\{ v_{t}\right\} ;t_{0},T;\bar{q}_{0},%
\bar{p}_{0},\eta _{0}\right] =\int_{t_{0}}^{T}\ell \left( s;f_{s},v_{s};\bar{%
q}_{s},\bar{p}_{s},\eta _{s}\right) ds+g\left( \bar{q}_{T},\bar{p}_{T},\eta
_{T}\right) .
\end{equation}
Here $\ell $ is a function of time, the current control policy values, and
current state parameters. The function $g$, known as a target or bequest
function in control theory, is a function of the state parameters at
termination. We assume that both are continuous in their arguments.

The cost $J$ will vary from one experimental trial to another, and must be
thought of as a random variable depending on the measurement output. The aim
of this section is to evaluate the minimum average cost over all possible
control policies, which we denote as 
\begin{equation*}
S\left( t_{0},T;\bar{q}_{0},\bar{p}_{0},\eta _{0}\right) =\min_{\left\{
f_{t}\right\} ,\left\{ v_{t}\right\} }\mathbb{E}\left\{ J\left[ \left\{
f_{t}\right\} ,\left\{ v_{t}\right\} ;t_{0},T;\bar{q}_{0},\bar{p}_{0},\eta
_{0}\right] \right\} .
\end{equation*}

\subsection{Bellman Optimality Principle}

For simplicity, let us write $z\equiv \left( \bar{q},\bar{p},\eta \right) $
and $u=\left( f,v\right) $ and $S\equiv S\left( t_{0};z_{t_{0}}\right) $,
etc. Taking $t_{0}<t_{0}+\Delta t<T$, we have that 
\begin{equation*}
S\left( t_{0};z_{t_{0}}\right) =\min_{\left\{ f_{t}\right\} ,\left\{
v_{t}\right\} }\mathbb{E}\left\{ \int_{t_{0}}^{t_{0}+\Delta t}\ell \left(
s;u_{s};z_{s}\right) ds+J\left[ \left\{ u_{t}\right\} ;t_{0}+\Delta
t,T;z_{0}+\Delta z\right] \right\}
\end{equation*}
where $\Delta z=z_{t}-z_{t_{0}}$ is, of course the random change in the
state parameters from time $t_{0}$ to $t_{0}+\Delta t$. We have that 
\begin{equation*}
\int_{t_{0}}^{t_{0}+\Delta t}\ell \left( s;u_{s};z_{s}\right) ds=\ell \left(
t_{0},u_{t_{0}},z_{t_{0}}\right) \,\Delta t+o\left( \Delta t\right)
\end{equation*}
up to terms that are small of order in $\Delta t$. Likewise, assuming that $%
S $ will be sufficiently differentiable, 
\begin{eqnarray*}
&&S\left( t_{0}+\Delta t;z_{0}+\Delta z\right) \\
&=&S\left( t_{0};z_{0}\right) +\left. \frac{\partial S}{\partial t}\right|
_{0}\Delta t+\left. \frac{\partial S}{\partial z}\right| _{0}\Delta z+\frac{1%
}{2}\Delta z^{\prime }\left. \frac{\partial ^{2}S}{\partial z^{2}}\right|
_{0}\Delta z+o\left( \Delta t\right) \\
&=&S\left( t_{0};z_{0}\right) +\left. \frac{\partial S}{\partial t}\right|
_{0}\Delta t+\left. \frac{\partial S}{\partial \bar{q}}\right| _{0}\left( 
\frac{1}{m}\bar{p}+v_{t}\right) \Delta t+\left. \frac{\partial S}{\partial 
\bar{p}}\right| _{0}\left( -\hbar \mu \bar{q}+f_{t}\right) \Delta t \\
&&+\left. \frac{\partial S}{\partial \eta ^{\prime }}\right| _{0}\frac{d\eta
^{\prime }}{dt}\Delta t+\left. \frac{\partial S}{\partial \eta ^{\prime
\prime }}\right| _{0}\frac{d\eta ^{\prime \prime }}{dt}\Delta t \\
&&+\frac{1}{2}\left. \frac{\partial ^{2}S}{\partial \bar{q}^{2}}\right| _{0}%
\left[ 2\kappa C_{qq}^{2}+2\tilde{\kappa}C_{qp}^{2}\right] \Delta t+\frac{1}{%
2}\left. \frac{\partial ^{2}S}{\partial \bar{p}^{2}}\right| _{0}\left[
2\kappa C_{qp}+2\tilde{\kappa}C_{pp}\right] \Delta t \\
&&+\left. \frac{\partial ^{2}S}{\partial \bar{q}\partial \bar{p}}\right|
_{0}2\sqrt{\kappa \tilde{\kappa}}\left[ C_{qq}^{2}+C_{pp}^{2}\right]
C_{qp}\Delta t+o\left( \Delta t\right) .
\end{eqnarray*}
(On the right hand side, we are evaluating at $t_{0},\bar{q}_{0},\bar{p}%
_{0},\eta _{0}$.)

The Bellman principle of optimality \cite{Bellman}, see also \cite{Davis}
for instance, states that if $\left\{ u_{t}^{\ast }\right\} $ is an optimal
control policy exercised over the time interval $\left[ t_{0},T\right] $ for
a given start state at time $t_{0}$, then if we operated this policy up to
time $t_{0}+\Delta t$ then the remaining component of the policy will be
optimal for the control problem over $\left[ t_{0}+\Delta t,T\right] $ with
start state being the current (random) state at time $t_{0}+\Delta T$. If we
assume the existence of such an optimal policy, then, within the above
approximations, as $\Delta t\rightarrow 0^{+}$, we are lead to the partial
differential equation (Hamilton Jacobi Bellman equation, or just Bellman
equation) for $S=S\left( t;\bar{q},\bar{p},\eta \right) $

\begin{gather}
0=\frac{\partial S}{\partial t}+\mathcal{H}\left( t;\bar{q},\bar{p},\eta ;%
\frac{\partial S}{\partial \bar{q}},\frac{\partial S}{\partial \bar{p}}%
\right) +\frac{\partial S}{\partial \eta ^{\prime }}\frac{d\eta ^{\prime }}{%
dt}+\frac{\partial S}{\partial \eta ^{\prime \prime }}\frac{d\eta ^{\prime
\prime }}{dt}  \notag \\
+\frac{\partial ^{2}S}{\partial \bar{q}^{2}}\left[ \kappa C_{qq}^{2}+\tilde{%
\kappa}C_{qp}^{2}\right] +2\frac{\partial ^{2}S}{\partial \bar{q}\partial 
\bar{p}}\sqrt{\kappa \tilde{\kappa}}\left[ C_{qq}+C_{pp}\right] C_{qp}+\frac{%
\partial ^{2}S}{\partial \bar{p}^{2}}\left[ \kappa C_{qp}^{2}+\tilde{\kappa}%
C_{pp}^{2}\right]   \label{Bellman equation}
\end{gather}
where we introduce

\begin{equation*}
\mathcal{H}\left( t;\bar{q},\bar{p},\eta ;y_{q},y_{p}\right)
:=\min_{f,v}\left\{ y_{q}\left( \frac{1}{m}\bar{p}+v\right) +y_{p}\left(
-\hbar \mu \bar{q}+f\right) +\ell \left( t;f,v;\bar{q},\bar{p},\eta \right)
\right\} .
\end{equation*}
It should perhaps be stressed that the derivation of this equation is
entirely classical. The key feature of the Bellman equation is that the
minimum is now taken pointwise: that is we look for the optimal scalar
values $f,v$ at a single instant of time. The equation is to be solved
subject to the terminal condition $\lim_{t\rightarrow T^{-}}S\left( t,\bar{q}%
,\bar{p},\eta \right) =g\left( \bar{q},\bar{p},\eta \right) $. 

In principle, once a minimizing solution $f^{\ast }=f\left( t;\bar{q},\bar{p}%
,\eta \right) ,v^{\ast }=v^{\ast }\left( t;\bar{q},\bar{p},\eta \right) $ is
known, it may be used as a Markov control for closed loop feedback: that is,
the control policies are taken as these functions of the current state
parameters.

The Bellman equations arising in quantum feedback control have so far proved
to be highly nonlinear and prohibitively hard to solve as a rule. Our
equation $\left( \ref{Bellman equation}\right) $ is no exception, however,
the nonlinearities are in due to the $\eta $ variable. We remark that if we
assume that we start off in a state relaxed at the equilibrium value $\eta
=\eta _{\infty }$, then the coefficients of the $\eta ^{\prime },\eta
^{\prime \prime }$ derivatives vanish exactly, and we may take the
covariances $C_{qq}$, $C_{qp}$ and $C_{pp}$ at their relaxed value
determined from  $\left( \ref{covariances}\right) $ evaluated at the
asymptotic value $\eta _{\infty }$. As the relaxation time is typically
small, we may justify this for large times $T$ in comparison. This ignores
any $\eta $-transient contribution to the cost, but at least opens up the
possibility of solving the Bellman equation and finding optimal Markov
control policies. We give the fundamental class of interest, quadratic
performance criteria, next.

\subsection{Linear Quantum Stochastic Regulator}

We consider the following quadratic control problem not involving any costs
on the $\eta $ parameter. In particular, we make the assumption that the
starting state is an asymptotic state $\left( \eta =\eta _{\infty }\right) $
and so we ignore $\eta $ as a variable. We set $x=\left( \bar{q},\bar{p}%
\right) $ and $u=\left( f,v\right) $ and take the specific choices 
\begin{eqnarray*}
\ell \left( t,u,x\right)  &=&\frac{1}{2}x^{\prime }A_{t}x+\frac{1}{2}%
u^{\prime }E_{t}u, \\
g\left( x\right)  &=&\frac{1}{2}x^{\prime }Rx,
\end{eqnarray*}
where $A_{t},E_{t}$ and $R$ are $2\times 2$ symmetric matrices with $E_{t}$
being invertible. The free Heisenberg equations are linear and can be
written as $\dot{x}_{t}=F_{t}x_{t}+M_{t}u$. The control problem is now
essentially the same as the classical stochastic regulator \cite{Davis}. In
this case we introduce a dual variable $y$ to $x$ and obtain the $\mathcal{H}
$-function
\begin{eqnarray*}
\mathcal{H}\left( t,x,y\right)  &=&\min_{u}\left\{ \ell \left( t,u,x\right)
+y^{\prime }\left( F_{t}x+M_{t}u\right) \right\}  \\
&=&\frac{1}{2}x^{\prime }A_{t}x+y^{\prime }F_{t}x+\min_{u}\left\{ \frac{1}{2}%
u^{\prime }E_{t}u+y^{\prime }M_{t}u\right\} 
\end{eqnarray*}
with the minimum attained at 
\begin{equation*}
u^{\ast }=-E_{t}^{-1}M_{t}^{\prime }y
\end{equation*}
and we find 
\begin{equation*}
\mathcal{H}\left( t,x,y\right) =\frac{1}{2}x^{\prime }A_{t}x+y^{\prime
}F_{t}x-\frac{1}{2}y^{\prime }M_{t}E_{t}^{-1}M_{t}^{\prime }y
\end{equation*}
Seeking an $\eta $-independent solution, the Bellman equation $\left( \ref
{Bellman equation}\right) $ reduces to 
\begin{equation*}
0=\frac{\partial S}{\partial t}+\mathcal{H}\left( t,x,\nabla S\right) +\frac{%
1}{2}K_{ij}\frac{\partial ^{2}S}{\partial x_{i}\partial x_{j}}.
\end{equation*}
Here $K$ is the matrix of the second order coefficients in $\left( \ref
{Bellman equation}\right) $ and these will be determined by the covariances $%
\left( \ref{covariances}\right) $ determined at the asymptotic value $\eta
_{\infty }$. As is well known \cite{Davis}, the solution takes the form $%
S\left( t,x\right) =\dfrac{1}{2}x^{\prime }\Sigma _{t}x+a_{t}$ where $\Sigma
_{t}$ satisfies the matrix Riccati equation 
\begin{equation*}
\frac{d\Sigma _{t}}{dt}=-\Sigma _{t}F_{t}-F_{t}^{\prime }\Sigma _{t}+\Sigma
_{t}M_{t}E_{t}^{-1}M_{t}^{\prime }\Sigma _{t}-A_{t},\qquad \Sigma _{T}=R,
\end{equation*}
while $a_{t}$ satisfies 
\begin{equation*}
\frac{da_{t}}{dt}=-tr\left\{ K\Sigma _{t}\right\} ,\qquad a_{T}=0.
\end{equation*}

The optimal control policy is therefore given by 
\begin{equation*}
u^{\ast }\left( t,x\right) =-E_{t}^{-1}M_{t}^{\prime }\nabla
S=E_{t}^{-1}M_{t}^{\prime }\Sigma _{t}x.
\end{equation*}

\subsection{Commentary}

The sufficiency property of the extended coherent states means that the
results above are of importance to the corresponding filtering problem.
Indeed this allows us to implement a quantum analogue of the Kalman filter
for state estimation amongst the class of extended coherent states. The
Kalman filter is of considerable conceptual and practical importance in
classical control theory and plays a crucial role in optimal feedback
control. In fact, the matrix Riccati equation occurring in linear stochastic
regulator also appears in a dual formulation as a Kalman filtering problem 
\cite{Davis}. Unfortunately, the solution to the fully parameterized Bellman
equation, that is, when we do not start from the equilibrium value $\eta
=\eta _{\infty }$, seems to be disappointingly difficult even in the linear
regulator example as the matrix $K$ will be quartic in $\eta $. (Such
difficulties seem to be sadly the norm in applications to optimal quantum
control as a whole, so far.) The control problem is however tractable for
the class of relaxed coherent states and corresponds to the linear regulator
model for quadratic performance and this at least gives us some insight into
possible applications.

\bigskip 

\textbf{Acknowledgment:} The author is indebted to Luc Bouten for a critical
reading of the article.

\end{document}